\documentclass[onecolumn,nofootinbib,aps,prd,preprintnumbers,10pt,longbibliography,floatfix]{revtex4-2}
\usepackage{filecontents}
\usepackage{subcaption}
\usepackage{float}
\usepackage{graphicx}% Include figure files
\usepackage{dcolumn}% Align table columns on decimal point
\usepackage{bm}% bold math
\usepackage{amsmath}
\usepackage{dsfont}
\usepackage{siunitx}
\usepackage[english]{babel}
\usepackage{amsfonts}
\usepackage{braket,slashed,bm}
\usepackage{cancel}
\usepackage{standalone} 
\usepackage{tikz} 
\usepackage[normalem]{ulem}
\usepackage{hyperref}
\hypersetup{
    colorlinks=true,
    linkcolor=blue,
    filecolor=blue,      
    urlcolor=blue,
    anchorcolor=blue,
    citecolor=blue
}
\DeclareFontFamily{U}{matha}{\hyphenchar\font45}
\DeclareFontShape{U}{matha}{m}{n}{
      <5> <6> <7> <8> <9> <10> gen * matha
      <10.95> matha10 <12> <14.4> <17.28> <20.74> <24.88> matha12
      }{}
\DeclareSymbolFont{matha}{U}{matha}{m}{n}

\DeclareMathSymbol{\Lt}{3}{matha}{"CE}
\DeclareMathSymbol{\Gt}{3}{matha}{"CF}
\def\beq{\begin{equation}}
\def\eeq{\end{equation}}
\def\barr{\begin{array}}
\def\earr{\end{array}}
\def\dis{\displaystyle}

\newcommand{\be}{\begin{equation}}
\newcommand{\ee}{\end{equation}}
\newcommand{\bea}{\begin{eqnarray}}
\newcommand{\eea}{\end{eqnarray}}

\newcommand{\bi}{\begin{itemize}}
	\newcommand{\ei}{\end{itemize}}

\begin{document}

\title{Baryon number violation from confining New Physics}

\author{Mathew Thomas Arun}%
\affiliation{School of Physics, Indian Institute of Science Education and Research Thiruvananthapuram, Vithura, Kerala, 695551, India}%
\date{\today}% It is always \today, today,

\begin{abstract}
The detection of neutron-antineutron oscillation will be a discovery of fundamental importance in particle physics and cosmology. In models discussed widely in literature, the process is generated through heavy New Physics with weak, perturbative, coupling at the scale of the experiment. Acknowledging the fact that Nature has been quite evasive regarding the strength and scale of New Physics, we discuss a new mechanism, generated by confining New Physics, at $\sim 2$ GeV, resulting in low-energy baryon number violating effects. The mechanism predicts baryon number violating processes like neutron disappearance, neutron-neutron annihilation and neutron-anti neutron oscillation generated through the condensation of the linear moose.
\end{abstract}

\maketitle

Though Nature has eluded all our direct searches for New Physics (NP) so far, it has bestowed us with a few indirect hints. Most prominent among them is the observed baryon to photon ratio~\cite{Planck:2018vyg}. This result implies violation of baryon number, thus bringing in asymmetry in baryon-anti baryon content of the universe. While the proton decay operator is suppressed to very high scales, the neutron-antineutron ($n-\bar{n}$) oscillation can be generated from New Physics at much lower scales~\cite{Mohapatra:1980qe,Babu:2006xc,Babu:2006wz,Babu:2008rq,Babu:2014tra,Dev:2015uca,Grojean:2018fus} by the dimension-9 effective field theory operator $(udd)(udd)$. On matching the Hamiltonian with the observed oscillation strength at Super-Kamiokande~\cite{Super-Kamiokande:2020bov}, the cut-off scale of the operator gets constrained to $\gtrsim \mathcal{O}(500 \text{ TeV})$. Moreover, the identification of 11 possible events with an expected background of $9.3\pm 2.7$ with $0.37$ megaton-year exposure of Super-Kamiokande\cite{Super-Kamiokande:2020bov}, it is highly likely that, neutron-antineutron oscillation might get detected at Hyper-Kamiokande. In all the models, discussed so far in literature, low-energy confining New Physics effects on baryon number violation is largely ignored. 

Mirror neutrons with $\sim$ GeV masses are studied in this context previously~\cite{Berezhiani:2005hv,Berezhiani:2006je,Pokotilovski:2006gq,Berezhiani:2015afa,Berezhiani:2009ldq,Berezhiani:2020vbe}, but in a $\sim\mathcal{O}(\text{TeV})$ scale New Physics perspective. This New Physics, at a very heavy scale, is assumed to introduce baryon number violating dimension-9 operators that contain thee SM quarks and three mirror quarks. These interactions generate neutron ($n$) to mirror neutron ($n^{\prime}$) conversion and neutron-antineutron oscillations with strengths $\epsilon_{nn^{\prime}} \sim \frac{(10 \text{TeV})^5}{\Lambda^5} \times 10^{-15} \text{eV}$, and $\epsilon_{n-\overline{n}} \sim \epsilon_{nn^{\prime}} \frac{\mu}{\Lambda}$~\cite{Berezhiani:2020vbe} respectively. Here, $\mu$ and $\Lambda$ are the new Majorana and mass scales of the New Physics that mediates the interaction between SM neutron and mirror neutrons. Comparing with the neutron-antineutron oscillation data, for New Physics at $\Lambda \sim 10 \text{TeV}$, the Majorana scale has to be $\mu \sim 10^{-6} \text{GeV}$, which unfortunately brings in large hierarchy between the scales. Moreover, this model again assumes perturbative couplings for the New Physics so that an effective field theory (EFT) description remains relevant. 

Due to the lack of our understanding regarding the nature of New Physics, it is important to acknowledge the possibility of baryon number violation originating from confining effects, at GeV scale. Such confining QCD-like theories are studied \cite{Georgi:1985hf} in moose notation, in the context of chiral symmetry breaking, compositeness, 't Hooft anomaly matching, etc, 
and are shown to be very rich, despite the apparent simplicity of the approach. Though complex models can be formulated to study the baryon number violating effects, an uncluttered presentation of the mechanism is possible with a minimal design of the linear moose. In Standard Model, the chiral symmetry ($U(N_f)_L\times U(N_f)_R$) breaks to $U(1)_V \times SU(N_f)_V$, where $N_f$ is the number of flavour, due to QCD condensation. Similarly there could exist numerous mirror chiral symmetries ($U(N_{f^{\prime}})_L\times U(N_{f^{\prime}})_R$) which are assumed to break to $U(1)_{\overline{V}} \times SU(N_{f^{\prime}})_{\overline{V}}$. Like in Standard Model where $U(1)_V$ is the baryon number global group, for mirror chiral symmetries $U(1)_{\overline{V}}$ becomes their respective mirror baryon number global group~\cite{Blinnikov:1982eh,Blinnikov:1983gh}. Being Standard Model singlets, such mirror sectors can talk to SM only gravitationally. Our aim in this article, is to identify the new interaction of these mirror sectors, with themselves and with SM, that can generate baryon number violation through condensation at GeV scale.

\begin{figure}
\centering
\vspace {-2cm}
\begin{subfigure}{0.4\textwidth}
\centering
    \includegraphics[width=6cm,height=9cm]{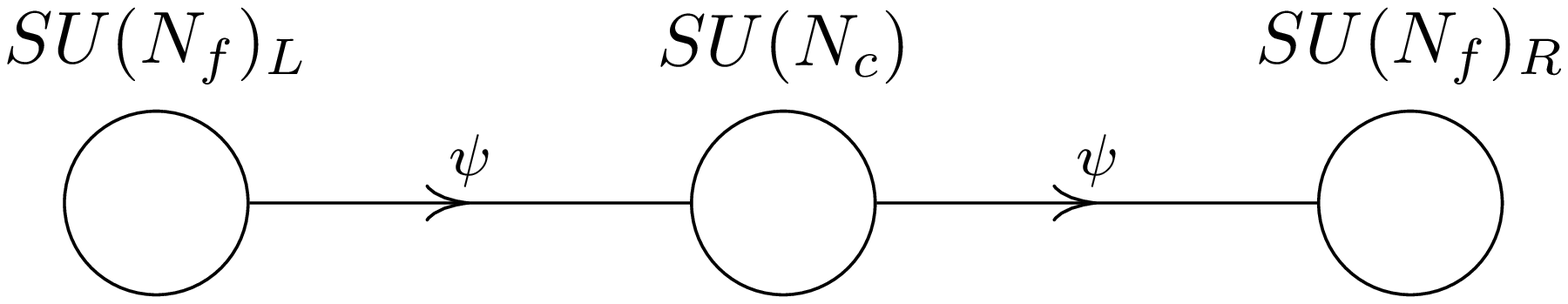}
\vspace {-3.5cm}
\caption{}
\label{fig:chssbsmall}
\end{subfigure}
%\hfill
\vspace {-2cm}
\begin{subfigure}{0.4\textwidth}
\centering
    \includegraphics[width=4cm,height=6cm]{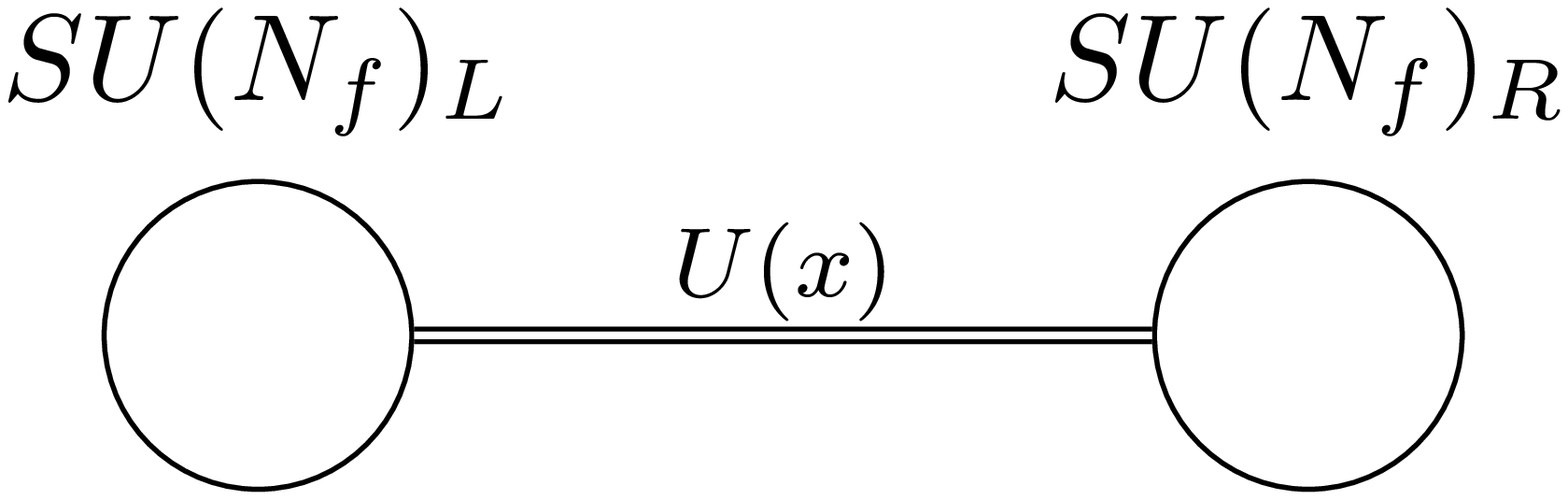}
\vspace {-2cm}
\caption{}
\label{fig:chssblarge}
\end{subfigure}

\vspace {2cm}
\caption{Figure (a) and Figure (b) shows the large energy and low energy behavior of QCD in linear moose language. The link fields with arrows are Weyl fermions, where as the double link represent the field whose vacuum expectation value breaks $SU(N_f)_L \times SU(N_f)_R$ to $SU(N_f)_V$.}
\label{fig:mirrormoose}
\end{figure}

To that end, lets first review the linear moose model discussed in the context of Standard Model chiral symmetry breaking~\cite{Georgi:1985hf}. In the linear moose language, thus, it is sufficient to represent the chiral symmetry breaking by $SU(N_f)_L \times SU(N_c)_c \times SU(N_f)_R$ at small distances. This is shown in Fig.\ref{fig:chssbsmall}. The fermion fields are represented at link fields with arrows such that arrow head flowing out and into the lattice site represent left and right handed Weyl-fermions respectively. Thus it is obvious that the $SU(N_c)_c$ gauge group does not posses anomaly. At large distances (low-energies), this gauge group confines and the chiral symmetry becomes $SU(N_f)_L\times SU(N_f)_R$. This is shown in Fig.\ref{fig:chssblarge}. The link fields are the aforementioned boson fields $U(x) = e^{\frac{2i}{f_\pi} \pi^a(x) T^a}$ whose vacuum, $\langle U \rangle =1$, spontaneously breaks the chiral symmetry to $SU(N_f)_V$. 

As an extension to this linear moose set up, lets assume the $U(1)_V$ and $U(1)_{\overline{V}}$ to be connected with each other through exotic new QCD like groups with confinement scale much greater than $\Lambda_{QCD}$. This model, at low-energies, will result in a lattice with each of the sites linked together by baryon number violating pseudo-Goldstone bosons of the condensation. For such NP models, it is impossible to express the neutron-antineutron oscillations as quark level effective operator. Instead, we need to study the dynamics at low energies in terms of free SM neutrons and antineutrons generated by the Lagrangian density~\cite{Berezhiani:2018xsx},
\begin{equation}
\mathcal{L} =  \bar{n}i\gamma^\mu \partial_\mu n - m_n \bar{n}n- \frac{\epsilon}{2} \Big[ \bar{n}^cn+\bar{n}\bar{n}^c\Big] \ ,
\label{eq:lagrangian}
\end{equation}
where $'n'$ is the four-component spinor field, identified as the SM neutron. The operator $\bar{n}^c n$ encompass the new physics contribution that generates $\Delta B=2$ hadronic transition matrix element,
\begin{equation}
\langle \bar{n}|\mathcal{H}_{\Delta B=2}|n\rangle = -\frac{1}{2}\epsilon \ \nu_{\bar{n}}^T C u_n \ , 
\end{equation}
where $\nu_{\bar{n}}$ and $u_n$ are Dirac spinors for $n^c$ and $n$ respectively.  

In this article, we discuss a non-perturbative baryon number violating mechanism that lead to baryon number violating  processes like neutron disappearance transitions, neutron-neutron annihilation and neutron-antineutron oscillations. While this mechanism successfully predicts an oscillation time period of $\tau_{n\bar{n}} \gtrsim 10^8 s$, naturally, bringing the confining New Physics within the reach of experiments~\cite{Chung:2002fx,Super-Kamiokande:2011idx,Bellerive:2016byv,Addazi:2020nlz,Ayres:2021zbh,Yiu:2022faw}, the neutron-mirror neutron conversion rate at Neutron Stars rule out beyond Standard Model states of mass $\lesssim 2 \text{GeV}$. The condensed linear moose also posses possible low-scale Baryogenesis upon the inclusion of CP violation and phase transition in the mirror sector.

Though this looks like a simple linear moose, the global chiral symmetry group of SM is assumed to be spontaneously broken via QCD condensation at the $U(1)$ lattice sites. Each of the $U(1)_{\overline{V}}$ lattice sites can inherit rich dynamics from their own QCD condensations. Since we are interested in the baryon charge and its violation, their internal symmetry mechanisms are not of much importance here. And for the current discussion, it is simpler to introduce the mirror neutrons and SM neutron as matter fields~\cite{Nakai:2014iea} located at the $U(1)$ lattice sites. We will briefly review the baryon in chiral perturbation theory for completeness in section\ref{sec:bchpt}. In section\ref{sec:toymodel} we will discuss a toy model with two lattice sites. We discuss the linear moose model and the generation of the baryon number violating currents in section\ref{sec:model}. The experimental constraints on the model are addressed in section\ref{sec:exptconst} and in section\ref{sec:UVmodel} we discuss a possible UV model. And finally we summarize the findings in section\ref{sec:conclusion}.

\section{Baryons in chiral perturbation theory: }
\label{sec:bchpt}

In this section, we briefly review the baryons in chiral perturbation theory. Lets consider a $SU(N_c)_c$ gauge theory coupled to fermions transforming under its fundamental representation. These fermions, are also assumed to be charged under the flavour global symmetry $U(N_f)_L \times U(N_f)_R$, such that
\begin{eqnarray}
U(N_f)_L &:& \dis \psi_{Li} \to L_{ij} \psi_{Lj} \nonumber \\
U(N_f)_R &:& \dis \psi_{Ri} \to R_{ij} \psi_{Rj} \ ,
\end{eqnarray}
where, $i,j={0,1,2,3,...,N_f}$ corresponding to the number of fermion flavours. Both $L$ and $R$ represent unitary matrices for the transformation. 

At low energies, with quarks forming chiral condensates, the only symmetry that remains unbroken is under the transformation for which $L=R$. Thus the chiral symmetry, $U(1)_V \times SU(N_f)_L \times SU(N_f)_R$, breaks to $U(1)_V\times SU(N_f)_V$, since the axial $U(1)_A$ group is anomalous. The Goldstone theorem now demands massless bosons corresponding to each of the broken generators. These excitations we can be parametrised as
\begin{equation}
U(x) = \xi^2(x) = e^{\frac{2i}{f_\pi} \pi^a(x) T^a} \ ,
\end{equation}
where the $\pi^a(x)$ are the meson fields and $f_\pi$ their decay constant. These fields transform as $U(x)\to L U(x)R^\dagger$ under the group. With $N_f=3$, this predicts the $N_f^2-1 = 8$ massless mesons. These are the pions, Kaons and the $\eta-meson$. The only obvious omission in this is the $\eta'-$meson, which is the would-be Goldstone boson associated with the anomalous $U(1)_A$ symmetry. Since this is broken by anomaly, in the chiral limit $\eta'-$meson is heavy and decouples from the low-energy effective action. 

On the other hand, though small, the non-vanishing quarks masses explicitly break the the chiral symmetry slightly. Incorporating this, the chiral Lagrangian for the pseudo-Goldstone bosons could be written as,
\begin{equation}
\mathcal{L}^{(2)}_{\pi \pi} = \int d^4x \frac{f_\pi^2}{4}\langle \partial^\mu U^\dagger \partial_\mu U \rangle + \frac{\sigma}{2} \langle mU + U^\dagger m^\dagger \rangle \ ,
\label{eq:chlagrangian}
\end{equation}
where $\langle ...\rangle$ denotes trace. Note that these pseudo-Goldstone bosons are neutral under the $U(1)_V$ symmetry. 

The symmetry at low-energies, also permit composite states of quarks that are charged under the unbroken $U(1)_V$ global group. These composite objects form the baryons, which are bosonic or fermionic depending on whether $N_c$ is even or odd respectively. Such modes are not present in the low-energy Lagrangian in Eq.\ref{eq:chlagrangian} and Fig.\ref{fig:chssblarge}. Instead, they appear as solitons in the chiral Lagrangian with a current,
\begin{equation}
J_B^\mu \propto \epsilon^{\mu \nu \alpha \beta} \langle L_\nu L_\alpha L_\beta \rangle \ ,
\end{equation}
where $L_\nu=(U^\dagger \partial_\nu U)$. This current is conserved obviously due to the antisymmetric property of the $\epsilon^{\mu \nu \alpha \beta}$ tensor and carries the winding number $\Pi_3 (SU(N_f)_V) = {\bf Z}$ given by the integral $B=\int d^3x J_B^0$. A low-energy model for these baryonic modes are given by the Skyrmion fields which are fermionic for $N_c=3$. Instead of working with Skyrme model, we can treat baryons as effective fermionic states that are vector like under $U(1)_V$. Since the gauge group ($SU(3)_c$) confines, they should be colour singlets formed by three quarks each charged $1/3$ under $U(1)_V$. From Weinberg-Witten theorem, such bound states should have helicity $\pm \frac{1}{2}$. Thus, at an effective hadronic level, the baryon number is realized as the symmetry under $U(1)_V$~\cite{Ecker:1994gg}.

The aforementioned fermions, given by,
\begin{equation}
\Psi = \frac{1}{\sqrt{2}}\begin{pmatrix}
\frac{1}{\sqrt{2}}\Sigma^0 +\frac{1}{\sqrt{6}}\Lambda & \Sigma^+ & p \\
\Sigma^- & -\frac{1}{\sqrt{2}}\Sigma^0 + \frac{1}{\sqrt{6}}\Lambda & n \\
\Xi^- & \Xi^0 & -\frac{2}{\sqrt{6}} \Lambda
\end{pmatrix} \ ,
\end{equation}
transform as $\Psi \to V(L,R,U(x)) \Psi $, where $V$ is a non-linear function of $L,R$ and the pseudo-Goldstone field $U(x)$, defined by the transformation $\xi \to L \xi V^\dagger = V \xi R^\dagger$, where $\xi = e^{i\frac{1}{f_\pi} \pi^a T^a}$. Now its straight forward to write down the chiral Lagrangian for baryons, to the leading order, as~\cite{Scherer:2002tk, Scherer:2005ri},
\begin{eqnarray}
\mathcal{L}_{eff} &=& \dis \mathcal{L}^{(2)}_{N\pi}  + \mathcal{L}^{(2)}_{\pi \pi} \ , \nonumber \\
\mathcal{L}^{(2)}_{N\pi} &=& \dis \bar{\Psi} \Big(i \gamma^\mu \partial_\mu - m + \frac{1}{2}g_{N\pi} \gamma^\mu \gamma_5 u_\mu\Big) \Psi
\label{eq:pionnucleonint}
\end{eqnarray}
where, $u_\mu = i \xi^\dagger \nabla_\mu U(x) \xi^\dagger$. In particular, the nucleon-pion interaction is given by,
\begin{equation}
\mathcal{L}^{(2)}_{N\pi} = \bar{\Psi} \Big(i \gamma^\mu \partial_\mu - m \Big)\Psi - \frac{g_{N\pi}}{2f_\pi} \bar{\Psi} \gamma^\mu \gamma_5 \Psi \partial_\mu \pi(x)^a T^a \ .
\label{eq:pionnucleonintexp}
\end{equation}
With this, we can build our moose model for the neutron-antineutron oscillation. But for introducing the philosophy and the mechanism, let's first discuss a toy model with two lattice sites. 

\section{Toy Model:}
\label{sec:toymodel}
\begin{figure}[H]
\centering
\vspace {-3cm}
    \includegraphics[width=11cm,height=14cm]{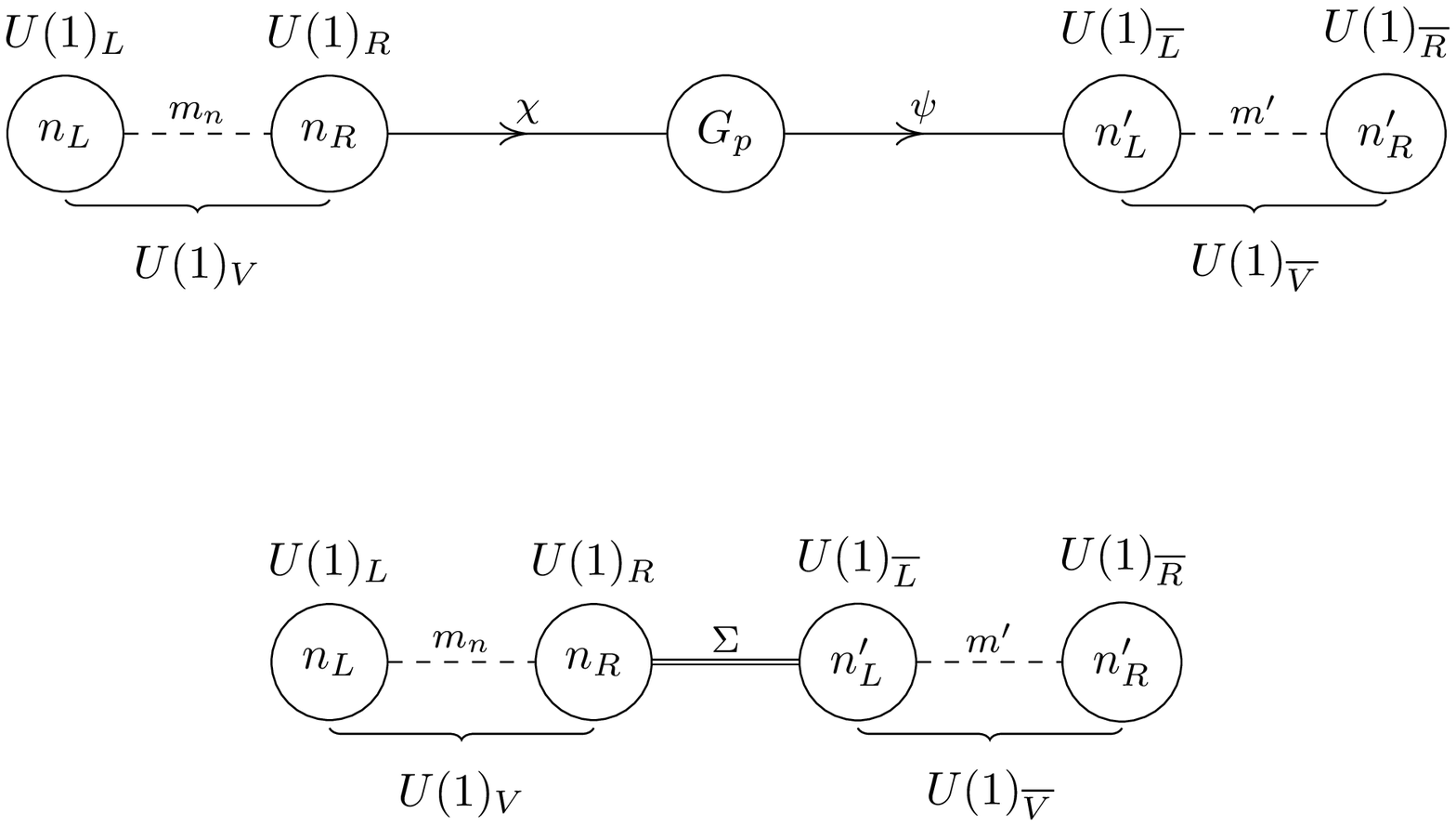}
\vspace {-4cm}
\caption{Lattice diagram of the two site toy model. At small distances (above) and after the strong group confines (below).}
\label{fig:twositemodel}
\end{figure}

Lets assume that there are two sets of chiral symmetries corresponding to SM and mirror sector. The full symmetry group is given by $[U(1)_V \times SU(N_f)_V] \times [U(1)_{\overline{V}} \times SU(N_{f^{\prime}})_{\overline{V}}]$ as shown in Fig.\ref{fig:twositemodel} linked together at the $U(1)$ sites. 

\begin{figure}[H]
\centering
\vspace {-6cm}
    \includegraphics[width=11cm,height=14cm]{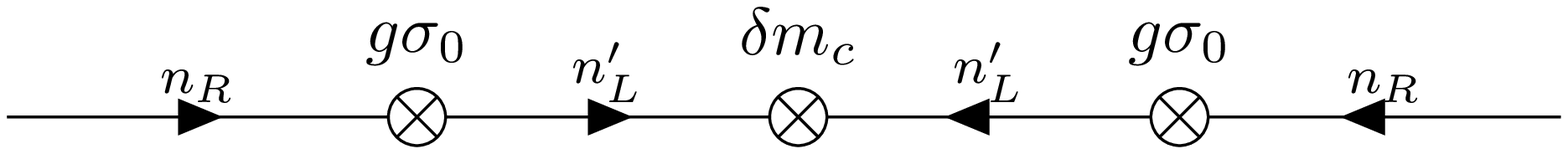}
\vspace {-6cm}
\caption{Neutron-antineutron oscillation mass insertion diagram for two site toy model}
\label{fig:massinstwositemodel}
\end{figure}

Standard Model baryon number group under which the SM baryons ($\Psi$) are charged is given by $U(1)_V$. Whereas, $U(1)_{\overline{V}}$ is the mirror baryon number global group, under which the mirror baryons (represented by $\Psi^{\prime}$) are charged. In the figure Fig.\ref{fig:twositemodel}, we have not shown the rest of the chiral group structure for simplicity. The chiral fermions, $\chi$ and $\psi$, are charged under these symmetry groups as $(B, p, 1)$ and $(1, p, \overline{B})$ respectively. With the gauge group $G_F$, in Fig.\ref{fig:twositemodel}, confining, the condensate $\langle \bar{\chi} \psi \rangle$ breaks the symmetry $U(1)_V \times U(1)_{\overline{V}}$ spontaneously to $U(1)_{V-\overline{V}}$. The new pions $\pi_p$, with charge $(-B, 1, \overline{B})$, mediate the interaction between the SM baryon ($\Psi$) and the mirror baryon $(\Psi^{\prime})$. Hence, the Lagrangian in Eq.\ref{eq:pionnucleonintexp} gets modified to,
\begin{eqnarray}
\mathcal{L}^{(2)}_{N\pi} &=& \dis \bar{\Psi} \Big(i \gamma^\mu \partial_\mu - m \Big)\Psi - \frac{g_{N\pi}}{2f_\pi} \bar{\Psi} \gamma^\mu \gamma_5 \Psi \partial_\mu \pi(x)^a T^a  \nonumber \\
&+& \dis \bar{\Psi}^{\prime} \Big(i \gamma^\mu \partial_\mu - m^{\prime} \Big)\Psi^{\prime} - \frac{g_{N^{\prime}\Pi}}{2f_\Pi} \bar{\Psi}^{\prime} \gamma^\mu \gamma_5 \Psi^{\prime} \partial_\mu \Pi(x)^a T^a \nonumber \\
&+& \dis g \bar{\Psi}^{\prime}_L \Sigma \Psi_R  + h.c. \ ,
\label{eq:pionnucleoninttoymodel}
\end{eqnarray}
where $\pi$ and $\Pi$ are the pions in the SM ($[U(1)_{V} \times SU(N_{f_N})_{V}]$) and the mirror ($[U(1)_{\overline{V}} \times SU(N_{f^{\prime}})_{\overline{V}}]$) chiral symmetries respectively. And the link field $\Sigma  = \sigma e^{i \frac{\pi_p}{f_p}} $. For the purpose of this work, lets only concern with the neutron ($n$) and its mirror partner ($n^{\prime}$). With the $\Sigma$ stabilizing to $\langle \Sigma \rangle = \sigma_0$, the quadratic part of the Lagrangian becomes,
\begin{eqnarray}
\mathcal{L}^{(2)} =  \mathcal{L}_{kinetic} -m_n \bar{n}_L n_R - m^{\prime} \bar{n}^{\prime}_L n^{\prime}_R  + g \sigma_0 \bar{n}^{\prime}_L n_R + h.c. \ .
\label{eq:nucleonintLagerangian}
\end{eqnarray}
This Lagrangian accommodates two Dirac neutrons that transform under the unbroken $U(1)_{V-\overline{V}}$ global group that preserve the $\Delta (B-\overline{B}) = 0$ number symmetry. Hence, if the mirror baryon number breaks, to preserve this symmetry the Standard Model baryon number has to break as well. The straight forward way to break the $U(1)_{\overline{V}}$ symmetry is to introduce Majorana terms in the mirror neutron sector. A Majorana term can can occur naturally if the chiral symmetry group in mirror sector is a real orthogonal group $SO(N_f)$~\cite{Georgi:1985hf} or can be induced via the vacuum expectation value of a scalar field charged under the mirror baryon symmetry~\cite{Berezhiani:2015afa}. In such cases the baryons are real and can posses Majorana and Dirac mass terms. Thus, introducing a term $\delta m_c \bar{n}^{\prime c} n^{\prime}$ breaks the mirror baryon number and in-turn breaks the baryon number by two units. For $g \sigma_0 \ll m^{\prime}$, using mass insertion, as shown in Fig.\ref{fig:massinstwositemodel}, to compute the strength of the baryon number violation, we get, 
\begin{equation}
\langle \bar{n}^c | n \rangle = \langle \bar{n}^c | n^{\prime c} \rangle \langle \bar{n}^{\prime} | n \rangle = \delta m_c \big(\frac{g \sigma_0}{m^{\prime}} \big)^2 \ .
\label{eq:neutronoscillationtwosite}
\end{equation}
This strength, from searches for neutron-antineutron oscillation at Super-Kamiokande~\cite{Super-Kamiokande:2011idx,Super-Kamiokande:2020bov}, should be $\sim 10^{-34} $ GeV. Since we are interested in low energy breaking of the baryon number, $m^{\prime} = m_n$, this brings unnatural smallness in the coupling or the Majorana scale.

On the other hand, for $g \sigma_0 \gtrsim m^{\prime}$, we need to minimize potential in Eq.\ref{eq:nucleonintLagerangian} with respect to $\bar{n}^{\prime}_L$. The equation of motion then becomes $n_R = \frac{m^{\prime}}{g \sigma_0} n^{\prime}_R$. Substituting this in the neutron mass term and introducing the mirror baryon symmetry breaking term $\delta m_c \bar{n}^{\prime c} n^{\prime}$, we get,
\begin{equation}
\mathcal{L} = -m_n \frac{m^{\prime}}{g \sigma_0}\bar{n}_L n^{\prime}_R + \delta m_c \bar{n}^{\prime c}_R n^{\prime}_R \ .
\end{equation}
This is a seesaw term that generates the baryon number breaking term, $m_n (\frac{m^{\prime}}{g \sigma_0}) \frac{1}{\delta m_c}$.

Both these scenarios require large fine-tuning of the characteristic scales in the model to satisfy the neutron-antineutron oscillations data. Instead, we will address this issue from a generalized perspective in the next section. 

%The second term in the above Lagrangian is troublesome. On renormalizing, the nucleon self-energy diagram would lead to corrections in the mass proportional to itself. This can be avoided if the chiral effective Lagrangian of these baryon modes can be developed by treating them as heavy static fields~\cite{Georgi,jenkins}. Assuming that the momentum transfer between baryons and pions are much smaller than the baryon mass. 

\section{Linear moose model:}
\label{sec:model}

Expanding on the toy model given in Fig.\ref{fig:twositemodel}, we will consider multiple QCD-like gauge groups with $U(1)_V\times SU(N_f)_L \times SU(N_f)_R$ chiral structure whose symmetry breaking leads to fermion modes in their respective low-energy effective Lagrangian. For the linear moose model, since we are interested in the baryon number violation, we need to consider only the various $U(1)$ global symmetry linked together as shown in Fig.\ref{fig:mirrormoose} with baryons and mirror baryons residing at these $U(1)$ sites. The link fields $\chi,\psi,f_1,g_2...$ are the new chiral fermions charged under new confining local gauge groups $G_p,G_s$ and should not be confused with the fermionic fields in mirror QCD sector. 

In this model, the $N+1$ mirror baryon lattice is generated from the condensation of $[U(1)_{\overline{V}}^{N+1} \times G_s^N]$ symmetry group, where $G_s$ is the strong gauge group with a confinement scale $\Lambda_s$. The $i^{th}$ lattice site, along with the fermions $f_i$ and $g_{i+1}$ (transforming as $(\overline{B}, s, 1)$ and $(1, s, \overline{B})$ under $[U(1)_{\overline{V}} \times G_s \times U(1)_{\overline{V}}]$), is shown in Fig.\ref{fig:mirrormoose}.

Interaction of SM baryon is introduced by linking the baryon symmetry group ($U(1)_{V}$) with $N=0$ sub-sector of the mirror neutron lattice through the gauge group $G_p$. The symmetry of this subsystem then becomes $[U(1)_{V} \times G_s \times U(1)_{\overline{V}}]$ connected through the fermion fields $\chi$ and $\psi$ with charges $(B,p,1)$ and $(1,p,\overline{B})$ under the subgroup. For simplicity and brevity, we assume $G_p = G_s \equiv SU(3)$.

\begin{figure}[H]
\centering
\vspace {-10cm}
    \includegraphics[width=18cm,height=24cm]{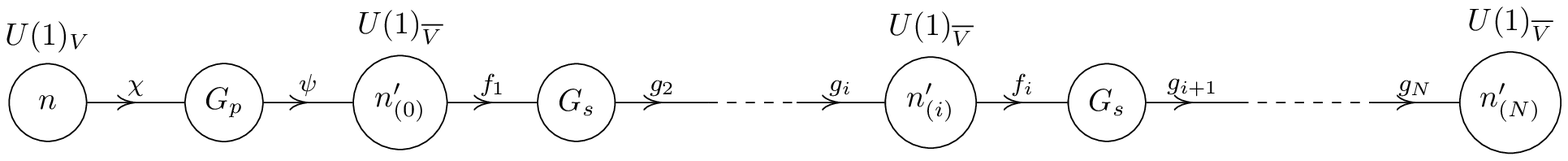}
\vspace {-10cm}
\caption{Lattice diagram of the model at short distances compared to the confinement scale $1/\Lambda_s$}
\label{fig:mirrormoose}
\end{figure}

\begin{figure}[H]
\centering
\vspace {-6cm}
    \includegraphics[width=11cm,height=15cm]{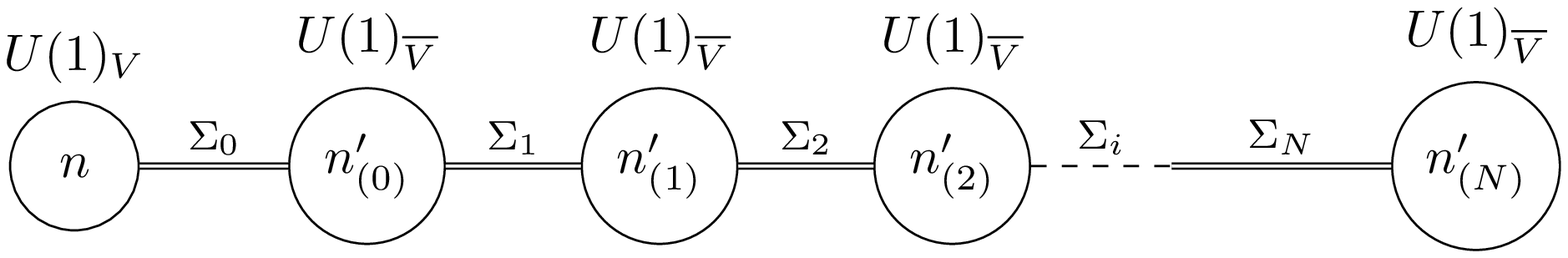}
\vspace {-6cm}
\caption{The figure shows the potential for neutron generated after the strong group confines.}
\label{fig:deconst}
\end{figure}

Thus, at short distances ($\ll 1/\Lambda_s$) the model possess $[U(1)_{\overline{V}} \times G_s]^{N+1} \times U(1)_{V}$ symmetry. The gauge groups condense at longer-distances ($\Lambda_s \sim \Lambda_{QCD}$), giving rise to the $[U(1)_{\overline{V}}]^{N+1}\times U(1)_{V}$ lattice linked together by the fermion condensates $\langle \bar{f}_i g_{i+1} \rangle = 4 \pi M^3 \Sigma_i$ and $\langle \bar{\chi} \psi \rangle = 4 \pi v^3 \Sigma_0$~\cite{Arkani-Hamed:2001kyx}, as shown in Fig.\ref{fig:deconst}, where $ v\equiv M = \Lambda_s/(4\pi)$. The low-energy effective action of the link fields ($\Sigma_i$, $i=0,1,2...N$) for this model becomes, 
\begin{eqnarray}
S&=& \dis \int d^4x \Big( M^2 \sum_{i=1}^N (\partial_\mu \Sigma_i)^\dagger \partial^\mu \Sigma_i + v^2 (\partial_\mu \Sigma_0 )^\dagger \partial^\mu \Sigma_0 \Big) + interactions\ ,
\end{eqnarray}
which can be brought to the canonical form by redefining $\Sigma_i \rightarrow M \Sigma_i$ and $\Sigma_0 \rightarrow v \Sigma_0$. 

With the vacuum expectation value of the condensate, $\langle \Sigma_0 \rangle$, spontaneously breaking the $[U(1)_{\overline{V}}]_0 \times U(1)_{V}$ symmetry to $U(1)_{V-\overline{V}}$, mirror baryons enter SM phenomenology to conserve the $B-\overline{B}$ number. Thus, $\Delta B \neq 0$ processes are rather susceptible to the effects of New Physics. In contrast to direct searches at hadronic colliders, precision measurements of neutron-antineutron oscillations~\cite{Berezhiani:2020vbe}, neutron decays at nuclei stability experiments~\cite{Phillips:2014fgb,Berezhiani:2018eds,Berezhiani:2018udo} and Neutron Star (NS) spin period and luminosity measurements~\cite{Berezhiani:2020zck,Goldman:2022brt} could be preferable to probe them.

The $U(1)_{V} \times U(1)_{\overline{V}} \rightarrow U(1)_{V-\overline{V}}$ breaking have other unique observable consequences as well. Though proton decay is forbidden for mirror neutrons heavier than SM neutron, elastic or inelastic conversion of SM neutron to mirror neutron in presence of Nucleons can be allowed. Along with lepton number violation, this will generate $pp\rightarrow e^+e^+$ (double proton annihilation) and Hydrogen-antiHydrogen oscillations.

In this condensed linear moose model, we describe the mirror and SM matter fields by two sets of Weyl spinors ($n^{\prime}_{(i)}$, $n^{\prime c}_{(i)}$) and ($n$, $n^{c}$) respectively located at N+2 sites of the lattice ~\cite{Nakai:2014iea} as shown in Fig.\ref{fig:deconst}. 
Low-energy phenomenology of this lattice is governed by the N+1 $\Sigma_i$ condensates and $\Sigma$. 
The condensate $\Sigma_i$ is charged $(-\overline{B},\overline{B})$ under the $[U(1)_{\overline{V}}^{i}\times U(1)_{\overline{V}}^{i-1}]$ subgroup while $\Sigma_0$ is charged $(-B,\overline{B})$ under $[U(1)_{V}\times U(1)_{\overline{V}}]$.
The effective Lagrangian for the system described by Fig.\ref{fig:deconst} is given by,
\begin{eqnarray}
\mathcal{L}^{(2)}_{N\pi} &=& \dis \bar{\Psi} \Big(i \gamma^\mu \partial_\mu - m_n \Big)\Psi - \frac{g_{N\pi}}{2f_\pi} \bar{\Psi} \gamma^\mu \gamma_5 \Psi \partial_\mu \pi^a(x) T^a  \nonumber \\
&+& \dis \sum_{i=0}^{N}\bar{\Psi}^{\prime}_{(i)} \Big(i \gamma^\mu \partial_\mu - m \Big)\Psi^{\prime}_{(i)} - \frac{g_{N^{\prime}\Pi}}{2f_\Pi} \bar{\Psi}^{\prime}_{(i)} \gamma^\mu \gamma_5 \Psi^{\prime}_{(i)} \partial_\mu \Pi^a_i(x) T^a \nonumber \\
&+& \dis  y \bar{\Psi}^{\prime}_{(0)L} \Sigma_0 \Psi_{R} + g \sum_{i=1}^{N}\bar{\Psi}^{\prime}_{(i)L} \Sigma_i \Psi_{(i-1)R}  + h.c. \ ,
\label{eq:pionnucleonintlinearmoosemodel}
\end{eqnarray}
where $\pi^a(x)$ and $\Pi_i^a(x)$ are the pions in the SM ($[U(1)_{V} \times SU(N_{f_N})_{V}]$) and $i^{th}$ mirror ($[U(1)_{\overline{V}} \times SU(N_{f^{\prime}})_{\overline{V}}]_i$) chiral symmetries respectively. And, $\Sigma_{i}= M e^{i \frac{\pi_{i}}{f_i}}, \ (i=0,1,2,...N)$ are the composite link fields connecting the lattice sites. In particular, $\Sigma_0 = v e^{i \frac{\pi_{0}}{f_0}}$, connects the SM neutron node $U(1)_V$ to the mirror neutron $n^{\prime}_{(0)}$ in $U(1)_{\overline{V}}$ node and rest of the $\Sigma_i's$ connect mirror lattice nodes as shown in Fig.\ref{fig:deconst}. Note that, in the above interaction term, the Yukawa coupling ($y$) is a spurion carrying the isospin charge. Ignoring the pion interactions and tabilising $\langle \Sigma_0 \rangle = v$ and $\langle \Sigma_{i} \rangle = M/g$, the neutron-mirror neutron Lagrangian that generates interactions in the lattice can be written as,
\begin{eqnarray}
\mathcal{L}^{(2)}_{N\pi} &\supset& \dis \bar{n}\Big(i\gamma^\mu \partial_\mu- m_n\Big)n + \sum_{i=0}^{N} \bar{n}^{\prime}_{(i)}\Big(i\gamma^\mu \partial_\mu - m \Big)n^{\prime}_{(i)}  +  y v \ \bar{n}^{\prime}_{(0)L}  n_R+ \sum_{i=1}^{N}\ M \bar{n}^{\prime}_{(i)L} n^{\prime}_{(i-1)R} \ \ .
\label{eq:mooselagrangian}
\end{eqnarray}
The potential felt by the SM neutron due to the presence of linear moose then becomes,
\begin{eqnarray}
\mathcal{V}_{n} &=& \dis yv \ \bar{n}^{\prime}_{(0)L} n_R+ m \ \bar{n}^{\prime}_{(0)L}n^{\prime}_{(0)R} + V \nonumber \\
\text{where, \ } V&=& \dis \sum_{i=1}^{N} M \ \bar{n}^{\prime}_{(i)L} n^{\prime}_{(i-1)R} -m \ \bar{n}^{\prime}_{(i)L} n^{\prime}_{(i)R} \ ,
\label{eq:moosepotential}
\end{eqnarray}
These terms in the above equation breaks the $ U(1)_{V} \times U(1)_{\overline{V}} \times ...  U(1)_{\overline{V}} $ symmetry to $U(1)_{V-\overline{V}}$, resulting in $B-\overline{B}$ number conservation.

Varying the potential with respect to $\bar{n}^{\prime}_{(i)L}$, and identifying the zeroth mirror mode ($n^{\prime}_{(0) R}$) with physical composite zero mode ($n^{\prime 0}_R$), we get,
\begin{equation}
n^{\prime}_{(i)R} = \Big(\frac{m}{M}\Big)^{i} n^{\prime 0}_R  \ .
\end{equation}
The diagonalisation introduces $\Delta B=1$ neutron to mirror neutron transition with strength $y v (\frac{m}{M})^N$ in Eq.\ref{eq:moosepotential}. Now, including an explicit breaking of mirror baryon number localised at the $N^{th}$ node of the moose the potential in Eq.\ref{eq:moosepotential} can be re-written as, 
\begin{eqnarray}
\mathcal{V}_{\cancel{\overline{B}}} &=& \dis \mathcal{V}_n + \frac{m_M}{2} \bar{n}^{\prime}_{(N)R} n^{\prime c}_{(N)R}  \nonumber \\
&=& \dis yv \ \bar{n}^{\prime}_{(0)L} n_R+ m  \ \bar{n}^{\prime}_{(0)L} n^{\prime 0}_{R} + \frac{m_M}{2}\Big(\frac{m}{M}\Big)^{2N} \bar{n}^{\prime 0}_{R} n^{\prime 0\ c}_{R}
\ .
\label{eq:potentialmbb}
\end{eqnarray}
The term that violates mirror baryon number can be generated with real representation of fermions transforming under the $SO(M)$ symmetry group. \footnote{The extra-dimensional interpretation would be to include real fermion mass term localised at the brane. In principle, this term can be located anywhere in the lattice, but a good extra-dimensional description may not be available always. }

The potential in Eq.\ref{eq:potentialmbb} now generates neutron-antineutron oscillation through an inverse seesaw like mixing, given by,
\begin{equation}
\mathcal{V}_{\cancel{\overline{B}}} \supset \frac{m_M}{2}\Big(\frac{m}{M}\Big)^{2N} \frac{y^2v^2}{m^2 + y^2v^2} \bar{n}_R n^c_R \ .
\label{eq:mixing}
\end{equation}
The Lagrangian density of SM neutrons now becomes,
\begin{eqnarray}
\mathcal{L}^{(2)}_{N\pi} &\supset& \dis \bar{n}\Big(i\gamma^\mu \partial_\mu  - m_n \Big)n -  \frac{m_M}{2}\Big(\frac{m}{M}\Big)^{2N} \frac{y^2v^2}{m^2 + y^2v^2} \bar{n}_R n^c_R + \mathcal{L}_{int} \ .
\end{eqnarray}
Comparing the above Lagrangian with Eq.\ref{eq:lagrangian}, we can see that the strength of neutron-antineutron oscillation as determined in the model becomes,
\begin{eqnarray}
 \frac{\epsilon}{2} &=& \dis  \frac{m_M}{2}\Big(\frac{m}{M}\Big)^{2N} \frac{y^2v^2}{m^2 + y^2v^2} \ .
\label{eq:nnbarosclinearmoose}
\end{eqnarray}
With Yukawa couplings of $\mathcal{O}(1)$, $m\sim yv$, $\frac{m}{M}\sim 0.1$, and $m_M \sim \mathcal{O}(1 \ \text{GeV})$, we get $\epsilon \sim 10^{-34} $ GeV for $\sim \mathcal{O}(15)$ mirror sites. 
This result translates to neutron-antineutron oscillation time period of $10^8 s$, which is within the reach of the current and future experiments~\cite{Phillips:2014fgb}. Note that the potential generates the necessary neutron-antineutron oscillation term without suppressed Yukawa. Thus, the constraints from neutron-mirror neutron transitions will be significant.

On the other hand, for $M\ll m$, the potential along the baryon number violating term leads to a generalised version of Fig.\ref{fig:massinstwositemodel} as shown in Fig.\ref{fig:moosemassins}. The strength of the neutron-antineutron oscillation is then given by $\frac{\epsilon}{2} = (yv)^2 m_M (\frac{M}{m})^{2N}$. 
\begin{figure}[H]
\centering
\vspace {-10cm}
    \includegraphics[width=15cm,height=22cm]{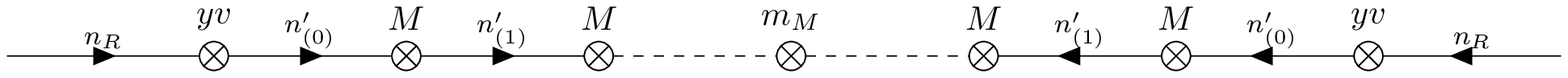}
\vspace {-10cm}
\caption{Neutron-antineutron oscillation mass insertion diagram for linear moose model.}
\label{fig:moosemassins}
\end{figure}

\section{Experimental constraints:}
\label{sec:exptconst}
\begin{figure}[H]
\centering
\vspace {-1cm}
    \includegraphics[width=5cm,height=8cm]{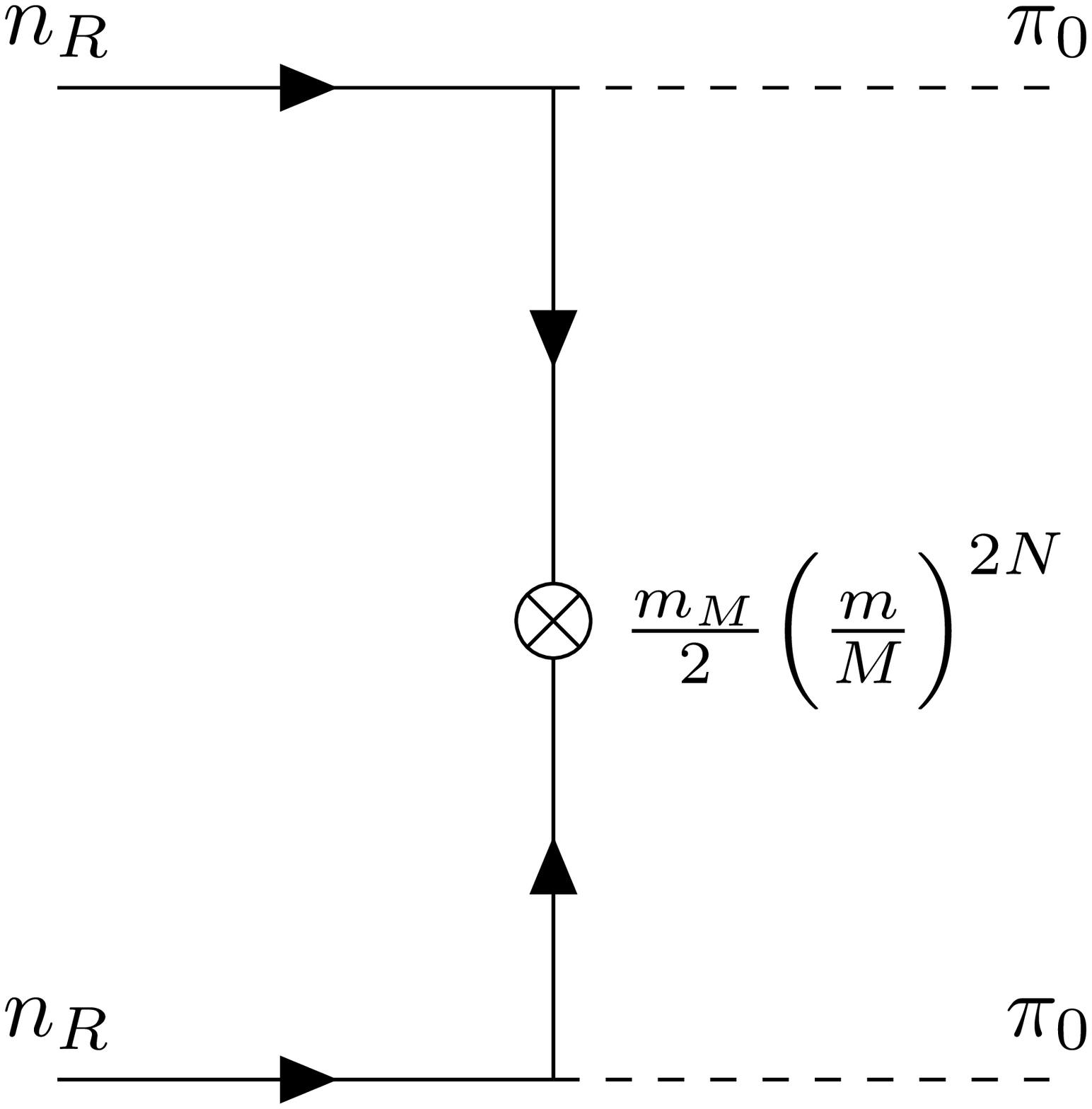}
\vspace {-1cm}
\caption{This figure shows the neutron-neutron annihilation mechanism through the effective inverse seesaw terms.}
\label{fig:moosemassins}
\label{fig:nntopipi}
\end{figure}

The dominant constraints on the model arise from the SM neutron to mirror neutron transition~\cite{Ayres:2021zbh} and baryon number violating annihilation of two SM neutrons to two $\pi_0$s. While the first process is generated by the direct mixing in Eq.\ref{eq:potentialmbb}, the second one proceeds through the tree level Feynman diagram as shown in Fig.\ref{fig:nntopipi}. 

Since the inverse seesaw mechanism, occurs naturally with unsuppressed Yukawa coupling, it is ruled out directly by the luminosity study of the Neutron Star~\cite{nEDM:2020ekj,McKeen:2021jbh}. The elastic SM neutron converting to mirror neutron in presence of a Nucleon ($\mathcal{N} \ n \rightarrow \mathcal{N} \ (n\rightarrow n^0)$), leaves a hole behind in the Fermi sea. This is refilled quickly by the neutrons with higher energy, and thus releasing the excess energy that can be measured in the luminosity~\cite{Guillot:2019ugf}. This constraint, on the other hand, relaxes for mirror neutron states with mass $\gtrsim 2 $ GeV, since enough energy is not available in the Neutron Star for the conversion. 

Nevertheless, double neutron annihilation ($nn \rightarrow \pi_0 \pi_0$), shown in  Fig.\ref{fig:nntopipi}, can occur. The exotic pions that are charged under baryon number, escapes the NS while leaving behind two holes, which again leads to the release of excess energy. This baryon number violating process is mediated by the mirror neutrons in the inverse seesaw potential, and the amplitude for the processes is suppressed by $(yv)^2 m_M (\frac{M}{m})^{2N}$. Though the process needs to be studied in detail, the Neutron Star can decay unless these exotic pions are also massive ($\gtrsim 2 $ GeV). Moreover, in~\cite{McKeen:2018xwc} it was noted that mirror neutron states with mass $\lesssim 2 $ GeV will soften the equation of state of Neutron Stars by reducing the Fermi pressure. Thus significantly affecting the stability of observed massive Neutron Stars.

Since in the previously mentioned process the final legs are SM singlets, it will also contribute to the dineutron invisible decay amounting to instability of the nucleus. Such processes are studied by the analyzing the energy deposition at the detectors arising from the deexcitation of the excited nuclei after $nn$ decays. In particular, at the $SNO+$ detector, this process leads to gamma rays in $5-10$ MeV range~\cite{SNO:2022trz}. The current partial lifetime limit on $nn \to invisible$ is $> 1.5 \times 10^{28} $ years. Though this is lesser than the neutron-anti neutron oscillation time period ($T_{n-\bar{n}} > 1.89 \times 10^{32}$) years in nucleus, it will strongly constraint the linear moose model. On the other hand, the model is safe from both Neutron Star and $SNO+$ measurements if these exotic states are heavier than $2$ GeV. This is also motivated because states much heavier than SM neutron do not take part in Big Bang nucleosynthesis.

\section{Ultra-Violet model:}
\label{sec:UVmodel}
Though in this article we successfully discussed the neutron-antineutron oscillation generated from confinement of strong gauge groups at low energies, a complete analysis of the interaction at the quark-level requires assuming an UV model. In this section, we will address the origin of the SM neutron-mirror neutron and mirror neutron-mirror neutron effective interactions, given in the last line in Eq.\ref{eq:pionnucleonintlinearmoosemodel}, from a quark-level New Physics perspective. 
First lets consider SM neutron-mirror neutron interaction operator and include few additional heavier scalar and fermion fields that transform under the gauge symmetries as given in table.\ref{table1}. 
\begin{table}[h]
  \begin{center}
    \begin{tabular}{|c|c|c|c|c|c|c|c|}
\hline
 & SM QCD & $G_p \equiv G_s$ & mirror QCD& & & &\\
      Field & $SU(3)_c$ & $  SU(3)_{p}$ & $SU(3)_{c^{\prime}}$& $U(1)_{Y}$& $SU(2)_{w}$& $B$ & $\bar{B}$\\
      \hline
	 & & & & & & &\\
	$\phi$ & $\overline{3}$ &1&1 & $\frac{2}{3}$ &1 & $\frac{2}{3}$&0\\
        $\phi^{\prime}$ & 1 & 3 & 1& 0&1&0&0\\
	$\omega^{\prime}$ & 1 &1& $\overline{3}$& 0&1&0& $\frac{2}{3}$\\
        $\xi_1 $ & 1 & $1$ & $1$& $0$&1 & 1&0\\
        $\xi_2 $ & 1 & $1$ & $1$& $0$&1&0&1\\
	$u_R$ and $d_R$ &3&1&1& $\frac{4}{3}$ and $-\frac{2}{3}$&1& $\frac{1}{3}$&0\\
	$u_{(0)}^{\prime}$ and $d_{(0)}^{\prime}$ &1&1&3& 0&1&0& $\frac{1}{3}$\\
	$\chi$ and $\psi$ &1&3&1& 0&1&1 and 0 & 0 and 1\\
\hline
    \end{tabular}
\caption{$SU(3)_c, G_p \equiv SU(3)_{p}$ and $SU(3)_{c^{\prime}}$ are the gauge groups of SM QCD, link fields $G_p\equiv G_s$ and the mirror QCD. The charges of the scalar fields, additional singlet fermions ($\xi_1$ and $\xi_2$), Standard Model quarks ($u$ and $d$), mirror fermions ($u^{\prime}$ and $d^{\prime}$) and link fermions($\chi$ and $\psi$) under these gauge groups are given above. $U(1)_Y$ and $SU(2)_w$ are the Standard Model hypercharge and weak gauge groups. Moreover, B and $\bar{B}$ denote the SM and mirror baryon number charges.}
\label{table1}
  \end{center}
\end{table}

\begin{figure}[h!]
\centering
\vspace{-2cm}
\includegraphics[width=9cm,height=9cm]{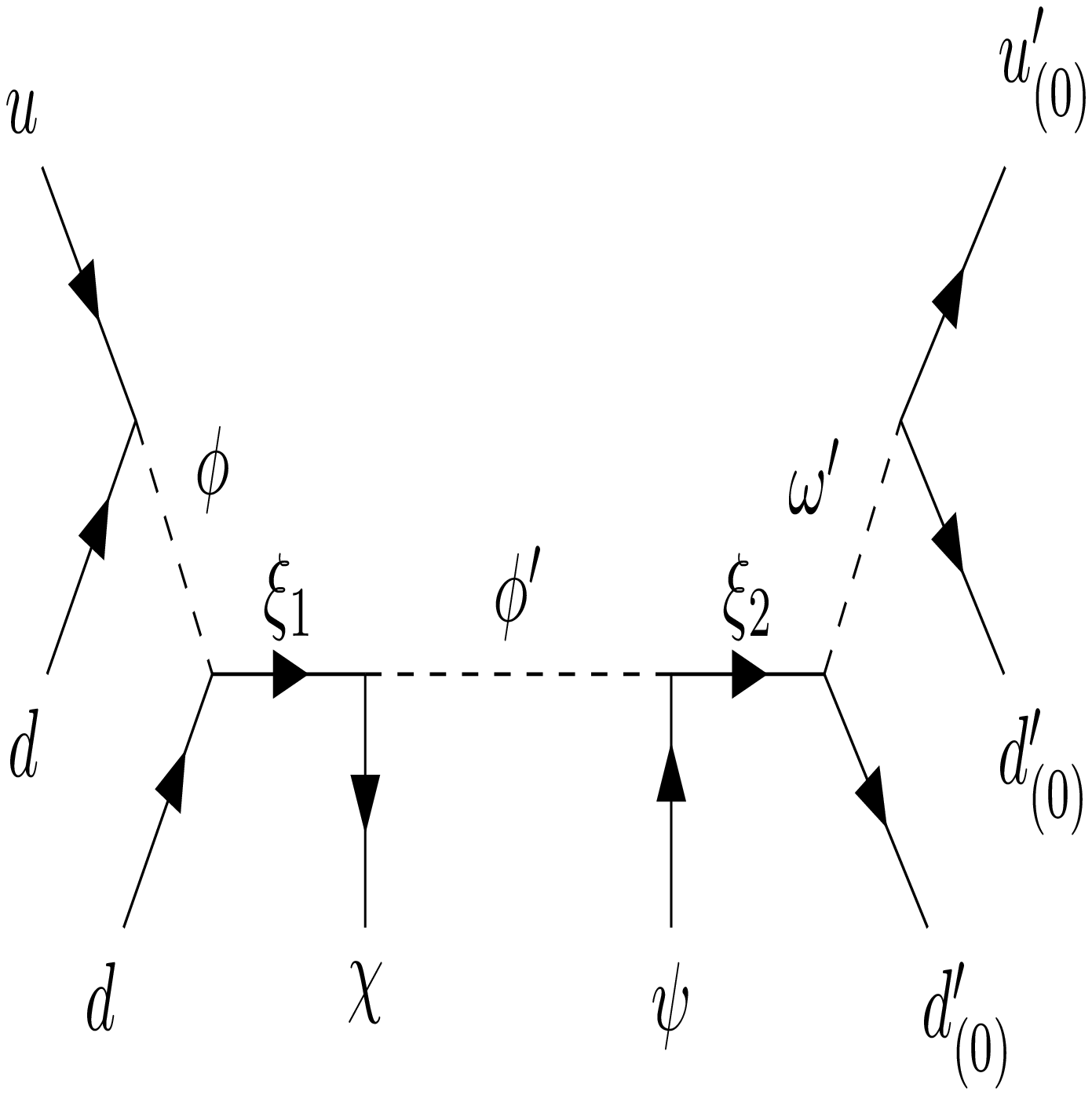}
\vspace{-2cm}
\caption{Feynman diagram showing the quark level interaction that leads to the SM neutron-mirror neutron effective vertex in Eq.\ref{eq:pionnucleonintlinearmoosemodel}, after the confinement of strong gauge groups.}
\label{fig:figure1}
\end{figure}

With these charge assignments, we can write the Lagrangian density for their interactions as,
\begin{equation}
\mathcal{L}_{int} = \lambda_1 \overline{u}^c d  \phi^\dagger + \lambda_2 \overline{\xi}_1 d \phi + \lambda_3  \overline{\chi} \xi_1 \phi^{\prime} + \lambda_4 \overline{\xi}_2 \psi \phi^{\prime \dagger} + \lambda_5 \overline{d}^{\prime} \xi_2 \omega^{\prime \dagger}  + \lambda_6 \overline{d}^{\prime} u^{\prime c} \omega^{\prime} ,
\end{equation} 
where, $u$ and $d$ are Standard Model quarks, while $u^{\prime}$ and $d^{\prime}$ denote mirror quarks that transform as fundamental under $SU(3)_{c^{\prime}}$. The fermion fields $\xi_1$ and $\xi_2$ are SM singlets but charged under the (SM, mirror)-baryon number as $(1,0)$ and $(0,1)$ respectively. This Lagrangian generates the interaction shown in Fig.\ref{fig:figure1}.

The effective interaction vertex of the SM neutron-mirror neutron, given in Eq.\ref{eq:pionnucleonintlinearmoosemodel}, is generated from Fig.\ref{fig:figure1} as,
\begin{eqnarray}
\mathcal{L} &=& \frac{\lambda_1\lambda_2\lambda_3\lambda_4\lambda_5\lambda_6}{m_{\phi}^2 m_{\phi^{\prime}}^2 m_{\xi}^2 m_{\omega^{\prime}}^2} \ \bar{\chi}\psi \ \overline{(u^{\prime} d^{\prime} d^{\prime})}_{(0)L}(u d d)_R \nonumber \\
&=& y  \ \bar{n}_{(0)L}^{\prime} \Sigma_0 n_R|_{\text{upon confinement}} \ ,
\end{eqnarray}
where $y \sim \frac{\lambda_1\lambda_2\lambda_3\lambda_4\lambda_5\lambda_6}{m_{\phi}^2 m_{\phi^{\prime}}^2   m_{\xi}^2 m_{\omega^{\prime}}^2 } (0.34 m_n)^3 m^{3} m_{\Sigma}^2$ is a dimensionless effective coupling, $m_n$ is the mass of the SM neutron, $m_{\Sigma}$ is the mass of the psuedo-Goldstone boson, $\Sigma_0 $, arising from the confinement of $G_p$ due to $\langle \bar{\chi} \psi \rangle$ condensate and $m$ is the mass of the mirror neutron. Moreover, we have denoted $m_\phi$, $m_{\phi^{\prime}}$ and $m_{\omega^{\prime}}$ for the masses of the three scalar fields and $m_{\xi}$ for the mass of the $\xi_1$ and $\xi_2$ fermion fields. 

Since, the scalar field $\phi$ carry $SU(3)_c$ charge, its mass is constrained strongly from $pp \to dijet$ resonant searches at LHC. The current lower bound on the mass of this scalar field is $8.2 \text{ TeV}$~\cite{ATLAS:2020iwa} for $\lambda_1\sim 1$ using run-2 $\sqrt{s}=13 \text{ TeV}$ with 139 $fb^{-1}$ data. On the other hand, the rest of the fields are singlets under Standard Model gauge group and hence are not constrained from colliders. Assuming that these fields are heavier ($\gtrsim \mathcal{O}(10 \text{ GeV})$) than the SM neutron, mirror-neutrons and confinement scales and couple with Yukawa interaction strengths $\lambda_1\sim 1$ and $\lambda_{2,3,4,5,6} \lesssim \mathcal{O}(4 \pi)$, the effective interaction coefficient becomes $y \sim 10^{-8}$.

Similarly, the mirror neutron interactions with other mirror neutrons at neighboring sites of the chain in Fig.\ref{fig:mirrormoose}, can be generated through the repetition of a similar operator described above with appropriately modified fields. These link couplings, given in Eq.\ref{eq:pionnucleonintlinearmoosemodel}, become, $g = \lambda^6 \frac{m^8}{\Lambda^8}$, where $\Lambda$ is the mediator mass. Since these fields are SM singlets, their masses are, again, not constrained by LHC but have to be heavier than $SU(3)_p, SU(3)_{c^{\prime}}$ confinement scales. Moreover, for the analysis here, like previously, lets assume their masses to be $ \Lambda \gtrsim \mathcal{O}(10 \text{ GeV})$. Using Eq.\ref{eq:nnbarosclinearmoose}, for $v = m \sim 2 \text{ GeV}$, $\frac{m}{M}\sim 0.1$ and $m_M \sim \mathcal{O}(10 \text{ GeV})$, the neutron-antineutron oscillation time period predicted by the model matches with the experimental limit for $N= 10$ sites in the linear moose chain. 

Other UV completion models can also be considered here and they may also satisfy the $n-\bar{n}$ experimental bound, for a given number of mirror sites, even though the mass of the QCD singlet New Physics fields are $\sim \mathcal{O}(few \text{ GeV})$. The confinement of strong groups $G_p \& G_s$ effectively increases the mass dimension of the neutron-antineutron oscillation operator and the expandability of the model to N sites efficiently incorporates low-scale New Physics contribution.

%in Eq.\ref{eq:mooselagrangian} and Eq.\ref{eq:moosepotential}

\section{Conclusion:}
\label{sec:conclusion}
In this article, we discuss a new mechanism for baryon number violation, generated from a simple linear moose with confining New Physics and mirror-baryon symmetries. At short distances this symmetry suppresses the mirror fermion interactions with SM. Precision low-energy experiments, on the other hand, prove to be excellent at probing the confinement effects of the New Physics. The spontaneous breaking of the mirror symmetry via condensation leads to $B-\overline{B}$ conserving interactions. Making baryon number violation measurements such as neutron-antineutron oscillation and Neutron Star luminosity measurements rather sensitive to the effect.

Though at Neutron Stars the effects of the linear moose can directly visible through $n + X_1 \rightarrow n^0 + X_2$ transitions, an explicit mirror symmetry breaking term is required in the mirror sector for $\Delta B=2$ transitions. Such terms are possible for real fermion representations of the internal chiral group. For heavier mirrors, neutron production by mirror neutron annihilations can lead to a possible low-scale Baryogenesis. This can be very interesting given the scenario accommodates phase transitions in the dark sector. The mirrors particles being heavier than the SM neutron makes sure that they do not affect the Big Bang Nucelosynthesis. 

In contrast to this non-perturbative mechanism, the $n-\bar{n}$ oscillation Hamiltonian, discussed widely in literature is generated by dimension-9 quark level operator suppressed by the energy scale $\Lambda \sim \mathcal{O}(500 \text{TeV})$. Such high scales are not achievable at direct search experiments for the foreseeable future. There has been efforts to bring the New Physics within the reach of current collider by selectively localising the quark wave profiles~\cite{Arkani-Hamed:1999ylh} and using orthogonal branes~\cite{Thomas:2022hyj} in extra-dimension models. Nevertheless, understanding of baryon number violation from non-perturbative behaviour of confining New Physics is very interesting and required.

\section*{acknowledgement}
M.T.A. acknowledges financial support of DST through INSPIRE Faculty grant [DST/INSPIRE/04/2019/002507]. The author thanks Debajyoti Choudhury for discussions. 

\bibliographystyle{unsrt}

\bibliography{biblio}

\end{document}